\documentclass[10pt,conference]{IEEEtran}

\usepackage{amsmath,amssymb,amsfonts}
\usepackage{cite}
\usepackage{graphicx}
\usepackage{textcomp}
\usepackage{multirow}
\usepackage{longtable}
\usepackage{lscape}
\usepackage{color, colortbl}
\usepackage{rotating}
\usepackage{wrapfig}
\usepackage{subcaption}
\usepackage{pdflscape}
\usepackage{hyperref}
\usepackage{array}
\usepackage{pifont}
\usepackage{balance}
\usepackage[most]{tcolorbox}
\usepackage{tikz}
\usepackage[normalem]{ulem}
\usepackage{url}
\usepackage{soul}
\usepackage{threeparttable}
\usepackage{xspace}
\usepackage{diagbox}
\usepackage{booktabs}
\usepackage{adjustbox}
\usepackage{svg}
\usepackage[ruled,vlined,lined,commentsnumbered]{algorithm2e}
\usepackage{amsmath,amsfonts}
\usepackage{enumitem}
\usepackage{fancybox}
\usepackage{fontenc}
\usepackage{listings}
\usepackage{longtable}
\usepackage{lscape}
\usepackage{makecell}
\usepackage{marvosym}
\usepackage{multicol}
\usepackage{multirow}
\usepackage{pifont}%
\usepackage{newtxmath}
\usepackage{rotating}
\usepackage{setspace}
\usepackage[most]{tcolorbox}
\usepackage{threeparttable}
\usepackage{tikz}
\usepackage[normalem]{ulem}
\usepackage{url}
\usepackage{soul}
\usepackage{wasysym}
\usepackage{indentfirst}
\usepackage{textcomp}
\usepackage{xcolor}
\usepackage{wrapfig}
\usepackage{pdflscape}
\usepackage{hyperref}
\usepackage{microtype}
\usepackage[skip=1pt,labelfont=bf]{caption}

\newcommand{\toolname}{RepoTransAgent\xspace}

\newcommand{\intuition}[1]{
\begin{tcolorbox}[colback=white,boxrule=1pt,top=0pt,bottom=0pt,left=1pt,right=2pt,top=2pt,bottom=2pt]
\em #1
\end{tcolorbox}
}

\begin{document}

\title{\toolname: Multi-Agent LLM Framework for Repository-Aware Code Translation}

\author{
\IEEEauthorblockN{Ziqi Guan$^{1}$\IEEEauthorrefmark{1}, Xin Yin$^{1}$\IEEEauthorrefmark{1}, Zhiyuan Peng$^{2}$, Chao Ni$^{1}$\IEEEauthorrefmark{2}}
\IEEEauthorblockA{
\textit{$^{1}$Zhejiang University, Hangzhou, China}, \{ziqiguan, xyin, chaoni\}@zju.edu.cn}
\IEEEauthorblockA{
\textit{$^{2}$Shanghai Jiao Tong University, Shanghai, China}, pzy2000@sjtu.edu.cn}
}

\maketitle

\begingroup
\renewcommand\thefootnote{\IEEEauthorrefmark{1}}
\footnotetext{
Equal contribution.}
\renewcommand\thefootnote{\IEEEauthorrefmark{2}}
\footnotetext{
Corresponding author.
}
\endgroup

\begin{abstract}
    Repository-aware code translation is critical for modernizing legacy systems, enhancing maintainability, and enabling interoperability across diverse programming languages. 
    While recent advances in large language models (LLMs) have improved code translation quality, existing approaches face significant challenges in practical scenarios: insufficient contextual understanding, inflexible prompt designs, and inadequate error correction mechanisms. 
    These limitations severely hinder accurate and efficient translation of complex, real-world code repositories.
            
    To address these challenges, we propose \toolname, a novel multi-agent LLM framework for repository-aware code translation. 
    \toolname systematically decomposes the translation process into specialized subtasks—context retrieval, dynamic prompt construction, and iterative code refinement—each handled by dedicated agents. 
    Our approach leverages retrieval-augmented generation (RAG) for contextual information gathering, employs adaptive prompts tailored to varying repository scenarios, and introduces a reflection-based mechanism for systematic error correction.
            
    We evaluate \toolname on hundreds of Java-C\# translation pairs from six popular open-source projects. 
    Experimental results demonstrate that \toolname significantly outperforms state-of-the-art baselines in both compile and pass rates. 
    Specifically, \toolname achieves up to 55.34\% compile rate and 45.84\% pass rate.
    Comprehensive analysis confirms the robustness and generalizability of \toolname across different LLMs, establishing its effectiveness for real-world repository-aware code translation.
\end{abstract}


\section{Introduction}

As software systems continue to evolve in scale and complexity, repository-aware code translation has emerged as a critical enabler for legacy system modernization and cross-platform development. Despite substantial advances in code translation research, existing approaches exhibit fundamental limitations when applied to repository-aware scenarios. Early approaches such as rule-based systems~\cite{nguyen2013lexical,chen2018tree} struggled with complex syntactic structures and lacked the adaptability required for diverse programming contexts, while Recurrent Neural Networks (RNN)-based models~\cite{dong2016language,yin2017syntactic,yin2018tranx} demonstrated limited effectiveness in handling repository-aware context. Furthermore, prompt designs employed in existing methods~\cite{khan2024xcodeeval,macedo2024exploring,pan2024lost,yang2024exploring} are typically simplistic and static, failing to adapt dynamically to varying repository contexts.

Recent Large Language Model (LLM)-based approa-ches~\cite{lano2024using,nitin2024spectra,pan2308understanding,yin2024rectifier} have demonstrated substantial improvements in translation quality, yet they remain constrained by insufficient contextual understanding and limited adaptability to real-world repository scenarios. Most prior work has concentrated primarily on isolated functions or small code snippets, systematically overlooking the global dependencies and complex invocation relationships that characterize large-scale software repositories.

Traditional static analysis techniques~\cite{ou2024repository,wang2024repotransbench}, including call graphs and program dependency graphs, pose significant interpretation challenges for LLMs and frequently introduce redundant or irrelevant information, thereby constraining their practical effectiveness. Furthermore, prompt designs employed in existing methods~\cite{khan2024xcodeeval,macedo2024exploring,pan2024lost,yang2024exploring} are typically simplistic and static, failing to adapt dynamically to varying repository contexts and consequently yielding suboptimal translation outcomes. Error correction strategies predominantly rely on test case execution results without providing actionable guidance or comprehensive context for effective issue resolution~\cite{pan2024lost,yang2024exploring}.

These fundamental limitations underscore the critical need for more flexible, context-aware, and adaptive solutions specifically tailored for repository-aware code translation. To address these challenges, we propose \toolname, a novel multi-agent LLM framework for repository-aware code translation. \toolname systematically decomposes the complex repository-aware translation task into specialized subtasks: retrieving semantically similar functions, collecting relevant contextual information, and performing iterative code generation and refinement. Each subtask is handled by a dedicated agent with domain-specific expertise.

The RAG Agent employs advanced retrieval-augmented generation techniques to identify and retrieve functions from the repository that exhibit semantic similarity to the source function. The Context Agent autonomously invokes specialized tools to incrementally gather necessary contextual information, focusing on the most relevant components for effective code translation. The Refine Agent performs the actual translation process based on the source function, target function signature, and comprehensive information collected by the RAG and Context Agents. Additionally, it executes test cases to collect execution feedback, reflects systematically on the output to generate targeted suggestions for error correction, and strategically re-invokes tools to acquire context specifically required for effective error resolution.

To ensure comprehensive agent understanding of current tasks and workflows, we design a dynamic and adaptive prompting scheme. The prompt architecture incorporates both static and dynamic components strategically. The static components include essential elements such as Goals, Guidelines, and Examples, which clarify task objectives, establish fundamental principles, and demonstrate basic workflows. The dynamic components encompass Gathered Context, Input, and Last Command, enabling agents to adapt their tool invocation strategies based on specific inputs and previously collected contextual information.

To evaluate the effectiveness of our approach comprehensively, we constructed a dataset of hundreds of Java-C\# translation pairs from six popular open-source projects on GitHub and conducted extensive experiments to assess \toolname's performance on repository-aware code translation tasks. The experimental results demonstrate that \toolname significantly outperforms baseline approaches in both compile and pass rates. Specifically, the average compile rate improved substantially from 26.07\%–30.47\% to 55.34\%, while the pass rate increased from 18.59\%–28.16\% to 45.84\%. Compared to baseline approaches, the functions successfully translated by \toolname almost completely encompass those translated by the three baseline methods while additionally achieving 90 more successful translations. We also conducted detailed analysis of error types for functions that \toolname failed to translate and found that most errors were compilation-related.

Our main contributions are summarized as follows:

\textbf{A. Multi-Agent Repository-Aware Translation Framework:} \toolname introduces a novel multi-agent LLM framework that systematically decomposes repository-aware code translation into specialized subtasks, enabling more accurate and context-aware translations.

\textbf{B. Dynamic Context Retrieval and Adaptive Prompting:} Our approach leverages advanced retrieval-augmented generation and dynamic prompt construction techniques, allowing agents to efficiently gather relevant context and adapt effectively to complex repository scenarios.

\textbf{C. Comprehensive Empirical Evaluation:} We conduct extensive experiments on real-world Java-C\# repositories, demonstrating that \toolname significantly outperforms state-of-the-art baseline approaches in both compile and pass rates.

\section{Motivation}

\subsection{Motivating Examples}
\begin{figure}[htbp!]
\centerline{\includegraphics[width=1\linewidth]{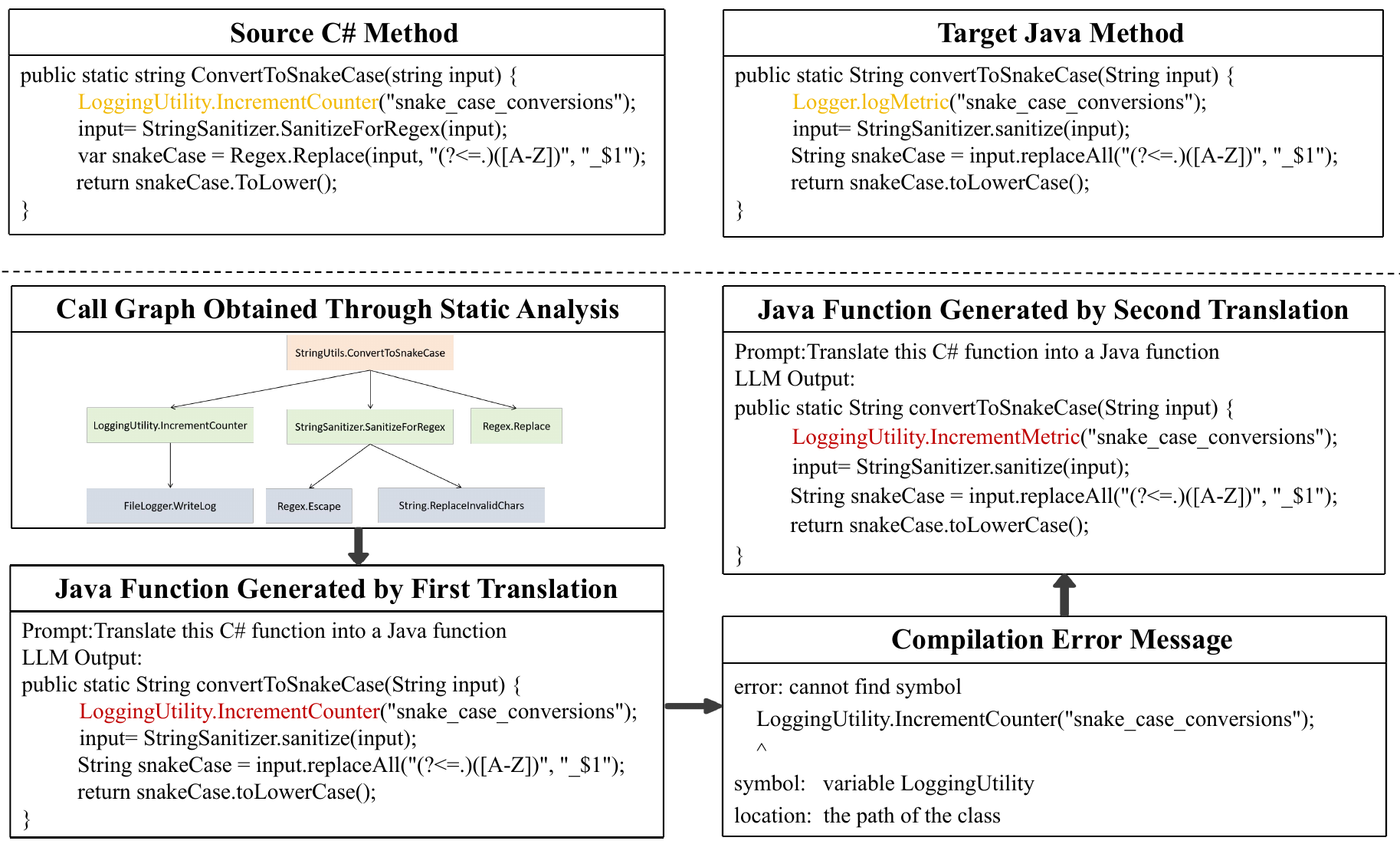}}
\caption{An example of repository-aware code translation using static analysis methods}
\label{fig:motivation_example_1}
\end{figure}

Consider the example illustrated in Fig.~\ref{fig:motivation_example_1}. The upper portion presents the C\# function to be translated and the corresponding ground truth Java function. Both functions perform camel case to snake case conversion while recording the number of conversions, yet they exhibit distinct logging utility implementations. The lower portion demonstrates the translation process of the C\# function using static analysis approaches.

\textbf{Challenge 1: Limitations of Static Analysis for Context Retrieval.}
Repository-aware code translation necessitates that LLMs access comprehensive contextual information to identify and understand dependencies, invocation relationships, and similar function implementations that provide translation guidance. Previous approaches~\cite{ou2024repository,wang2024repotransbench} typically employ static analysis techniques, such as call graphs and program dependency graphs, to collect necessary context. However, these graph-based representations are inherently complex and difficult to convert into formats readily interpretable by LLMs, significantly constraining their effectiveness. Moreover, static analysis methods rely on predefined rules and pattern matching, which prove inadequate for handling unforeseen code structures or dynamic changes in external dependencies. Static analysis can also introduce excessive and often irrelevant context, increasing the risk of LLM hallucination. These fundamental limitations severely constrain the performance of traditional static analysis-based code translation approaches.

\textbf{Challenge 2: Inadequacy of Fixed Prompt Designs.}
Previous approaches~\cite{yin2024rectifier,khan2024xcodeeval,macedo2024exploring,pan2024lost,yang2024exploring} predominantly rely on simplistic and fixed prompt designs. While such prompts offer implementation simplicity, their quality significantly impacts LLM effectiveness. These rigid prompts prove inadequate for the complex scenarios encountered in repository-aware code translation, failing to capture subtle structural and functional differences between functions and lacking adaptability to varying contexts. Consequently, they constrain the full potential of LLMs and frequently lead to erroneous translation results.

\textbf{Challenge 3: Insufficient Error Correction Mechanisms.}
Previous approaches~\cite{pan2024lost,yang2024exploring} typically rely on test case execution results to correct errors in translated code. However, these results lack comprehensive solutions or essential information required for effective error resolution. Simply providing erroneous code and related test failures proves insufficient for LLMs to successfully correct mistakes, resulting in ineffective repair attempts. For instance, error messages may only indicate that a class symbol cannot be found without providing resolution information, making it challenging for LLMs to perform effective corrections. In real-world scenarios, developers typically begin by understanding error root causes, gathering necessary context, and then proposing targeted fixes—a systematic process that yields more effective error correction.

\subsection{Key Ideas}
Based on these observations, we propose \toolname for repository-aware code translation with the following key innovations:

\textbf{Key Idea 1: Multi-Agent LLM Framework.}
Repository-aware code translation encompasses multiple complex subtasks, including retrieval-augmented generation (RAG), context collection, code generation, and iterative error correction based on test case execution results. Each subtask requires meticulous execution for overall task success. To address this complexity, we design a multi-agent framework where specialized agents handle distinct subtasks: the RAG Agent retrieves semantically similar functions, the Context Agent gathers necessary contextual information, and the Refine Agent manages code generation and iterative refinement. By delegating tasks to specialized agents that collaborate systematically, our approach enhances the effectiveness, maintainability, and robustness of repository-aware code translation.

\textbf{Key Idea 2: Intelligent Context Retrieval via Tool Invocation.}
Within the Context Agent, we design five distinct tool functions that enable selective invocation based on the agent's analytical requirements. The agent demonstrates intelligent behavior by identifying critical aspects of the target function relevant to code translation and efficiently retrieving essential contextual information. This approach minimizes redundant context and improves translation efficiency. Additionally, the agent possesses reasoning capabilities, enabling interpretation of natural language descriptions in code (such as comments and method names) and leveraging domain knowledge to infer semantic information. This leads to more precise context retrieval and addresses the inflexibility and redundancy issues inherent in traditional static analysis approaches. All tool outputs are returned in a unified JSON format, ensuring straightforward interpretability by LLMs and overcoming the complexity of traditional static analysis formats.

\textbf{Key Idea 3: Dynamic Prompt Construction.}
To overcome the limitations of traditional prompt designs—which are often simplistic and lack sufficient quality and relevance—we adopt dynamic and well-structured prompts in both the Context Agent and Refine Agent. Each prompt comprises clearly defined components: the Goals section specifies the agent's task objective, the Guidelines section outlines fundamental principles for task execution, the Example section demonstrates the task process through concrete illustrations, and the Output Format section defines required response formats. Additionally, we dynamically adjust prompt content based on the agent's current state. For instance, the Input section updates according to the source function being translated, while the Gathered Context section incorporates previously collected contextual information. This highly customized and structured prompt design significantly enhances LLM performance in complex repository-aware code translation tasks, fully exploiting their reasoning and generation capabilities.

\textbf{Key Idea 4: Enhanced Error Correction with Contextual Analysis.}
In the Refine Agent, after obtaining erroneous code and test case execution results, we require the agent to analyze error causes based on faulty code and test outputs, subsequently proposing targeted correction strategies. Additionally, we instruct the agent to reinvoke tool functions to gather context closely related to identified errors. This approach not only generates repair suggestions but also retrieves necessary contextual information for error resolution, significantly increasing the probability of successful correction and emulating the systematic debugging process employed by experienced developers.
\label{sec:motivation}

\section{Approach}

\subsection{Overview}

\begin{figure}[htbp]
\centerline{\includegraphics[width=1\linewidth]{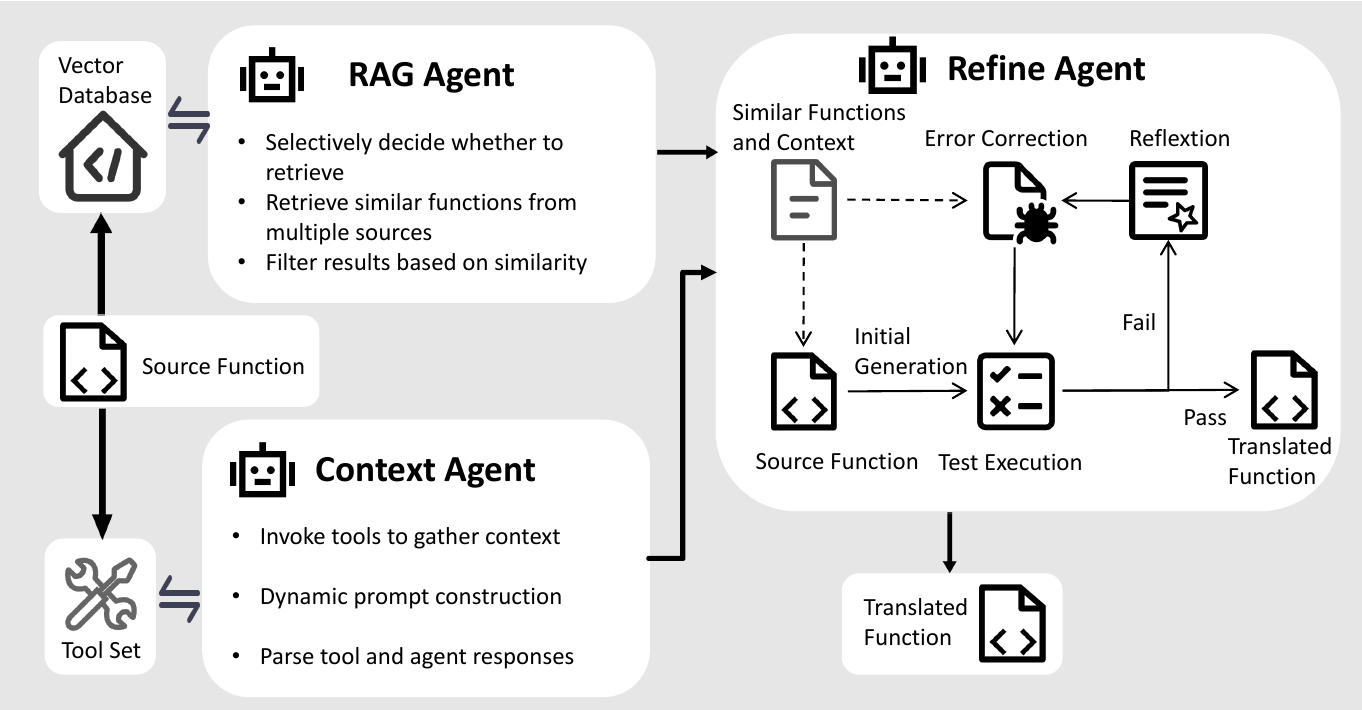}}
\caption{Overview of \toolname}
\label{fig:overview}
\end{figure}

    Fig.~\ref{fig:overview} provides an overview of the \toolname approach, which consists of three specialized agents: RAG Agent, Context Agent, and Refine Agent. During repository-aware code translation, these three agents work collaboratively to generate the final translated code. The RAG Agent is responsible for retrieving functions similar to the target function to assist in translating the source function. The Context Agent retrieves the contextual information required for translating the source function. The Refine Agent performs iterative refinement by translating the code, generating reflections based on test case execution results, and re-retrieving context according to the error information. It then uses the reflection, error messages, and updated context to fix the mistakes made in the previous iteration.

\subsection{RAG Agent}
This section details how we employ the RAG Agent to retrieve functions semantically similar to the target function to assist in translating the source function. The RAG Agent comprises four main components: Preprocessing, Selective Retrieval, Multi-Route Retrieval, and Result Filtering.

\subsubsection{Preprocessing}
To enable effective retrieval, we first store the project's metadata in a vector database. Given the metadata of a project repository, we begin by preprocessing it and generating embeddings for two distinct components of the metadata, which are subsequently stored in the vector database. The first component consists of the method bodies of all source-target translation pairs, and the second component includes the function names of all methods in the target repository. These two components are stored separately in two vector databases, referred to as the pair store and the name store, respectively.

\subsubsection{Selective Retrieval}
In real-world code repositories, many functions are standalone, meaning they do not reference external classes or functions. For such functions, it is unnecessary to employ the RAG process to retrieve additional information during translation. Bypassing the RAG process in these cases can conserve computational resources and prevent redundant information from adversely affecting translation quality. Therefore, before initiating the RAG process, we first enable the agent to determine whether the target function requires RAG assistance. This decision is based primarily on whether the target function is standalone and whether it exhibits sufficient simplicity. If the agent determines that RAG is not needed, the RAG process is bypassed.

\subsubsection{Multi-Route Retrieval}
If the agent determines that a target function requires the RAG process, a multi-route retrieval procedure is initiated. This procedure consists of two complementary components. The first component involves retrieving source-target translation pairs where the source function is similar to the current source function from the pair store. A hybrid retrieval strategy is employ: both dense retrieval (based on cosine similarity) and sparse retrieval (based on BM25) are utilized to independently retrieve the top-k most similar results from the pair store. These results are subsequently ranked using the Reciprocal Rank Fusion (RRF) method to produce the final top-k most relevant results. The second component involves retrieving functions from the name store based on the target function's name. Using sparse retrieval (BM25), the top-k functions with the most similar names are retrieved. If a function name exactly matches the target function name, its similarity score is set to 1.0. Finally, the results from both retrieval paths are merged to form the output of the multi-route retrieval step.

\subsubsection{Result Filtering}
Not every function retrieved through the multi-route retrieval process is beneficial for translating the source function into the target function. Including irrelevant functions as additional context can be not only unhelpful but potentially detrimental to translation quality. Therefore, it is essential to filter the retrieved results to retain only those functions that are genuinely useful.

We leverage the agent to perform this filtering operation. For each candidate function in the retrieved set, the agent compares it with the target function, considering factors such as functional similarity, structural resemblance, and referenced context. Based on this comparative analysis, the agent determines whether the candidate function is sufficiently similar to the target function to provide useful guidance for translating the source function. Functions deemed relevant are retained, while others are discarded.

\subsection{Context Agent}
This section presents how we employ the Context Agent to obtain the necessary context for repository-aware code translation. It comprises three components: Workflow, Tool Introduction, and Dynamic Prompt Construction.

\subsubsection{Workflow}
In the workflow of the Context Agent, we first construct a dynamic prompt using relevant information from the source and target functions, and use this prompt to query the agent, instructing it to decide which tool to invoke. The agent's response strictly adheres to a JSON format and includes three parameters: id (the identifier for the tool invocation), name (the name of the tool to be utilized), and args (the arguments for the tool).

After obtaining the tool invocation command from the agent, the Context Agent proceeds to call the specified tool and retrieves the corresponding output. This output is added to the context collection, and we also log relevant information from the tool invocation. We then use this information from the previous round to reconstruct a dynamic prompt and prompt the agent to continue invoking tools. This process is repeated iteratively until either the maximum number of iterations is reached or the agent determines that sufficient context has been gathered. Finally, the context collection is returned as the context retrieved by the Context Agent.

\subsubsection{Tool Introduction}
Our approach introduces five tool functions designed to help the Context Agent acquire context most pertinent to the source function's translation. In the Context Agent's workflow, the agent autonomously decides which tools to invoke based on function information and historical interaction data, and utilizes the outputs of these tools as contextual information. The following is an introduction to each of these tools:

\paragraph{get\_source\_class\_info} 

get\_source\_class\_info returns all field definitions and method signatures within the class that contains the source function. This tool helps the agent deduce the class to which an instance belongs by combining information about class instances and fields from the source function. This facilitates the agent in exploring how classes referenced in the source function are implemented in the target repository.    

\paragraph{get\_target\_class\_info}

get\_target\_class\_info retrieves all field definitions and method signatures within the class that contains the target function. Since the target function is likely to reference these predefined fields and implemented methods, this tool helps the agent understand which fields or methods may be invoked within the target function.

\paragraph{find\_target\_imports} 

find\_target\_imports retrieves all the imported classes in the file that contains the target function. These imported classes may be referenced within the target function and also reveal the dependencies of the target class.

\paragraph{find\_target\_class\_info} 

Given a class name to be searched, find\_target\_class\_info searches all field definition information and method signature information of the class matching the class name in the target repository. If the agent believes that a certain class has a high probability of being referenced in the target function, this tool function can effectively help the agent find relevant information about the class. The agent can infer which classes may be imported in the target function based on the class name, or determine if the return value and parameters of the target function signature contain external classes. Additionally, if an instance of an external class is utilized in the source function, there is a high probability that the target function will also use an instance of that class. The agent uses get\_source\_class\_info to obtain the name of the class to which the instance belongs, and then uses find\_target\_class\_info to query the implementation of the class in the target repository. This approach addresses cases where the context of the class in the source repository and the target repository is inconsistent.

\paragraph{find\_target\_method\_body} 

Given a method name and its owning class name, the find\_target\_method\_body tool function searches in the target repository for the class that matches the class name, and then looks in that class for the complete contents of the method that matches the method name, including its method signature and method body. If a method is likely to be utilized in the target function, then understanding the specific logic and functionality of the method can help the agent better utilize the method. At the same time, knowing the specific function of a method allows the agent to decide whether to use the method in the target function.

In the workflow of the Context Agent, we first require the agent to prioritize the invocation of three tool functions: get\_source\_class\_info, get\_target\_class\_info, and find\_target\_imports, in order to collect fundamental information relevant to the translation of the source function. After obtaining this basic information, the agent gains a preliminary understanding of the context required for translating the source function. Subsequently, the agent can invoke the find\_target\_class\_info and find\_target\_method\_body tool functions multiple times to gather more specific information, continuing until it determines that a sufficient amount of context has been collected for the code translation. All context retrieved through these tool calls is finally consolidated into a JSON format and provided to the Refine Agent.

\subsubsection{Dynamic Prompt Construction}
\begin{figure}[htbp]
\centerline{\includegraphics[width=1\linewidth]{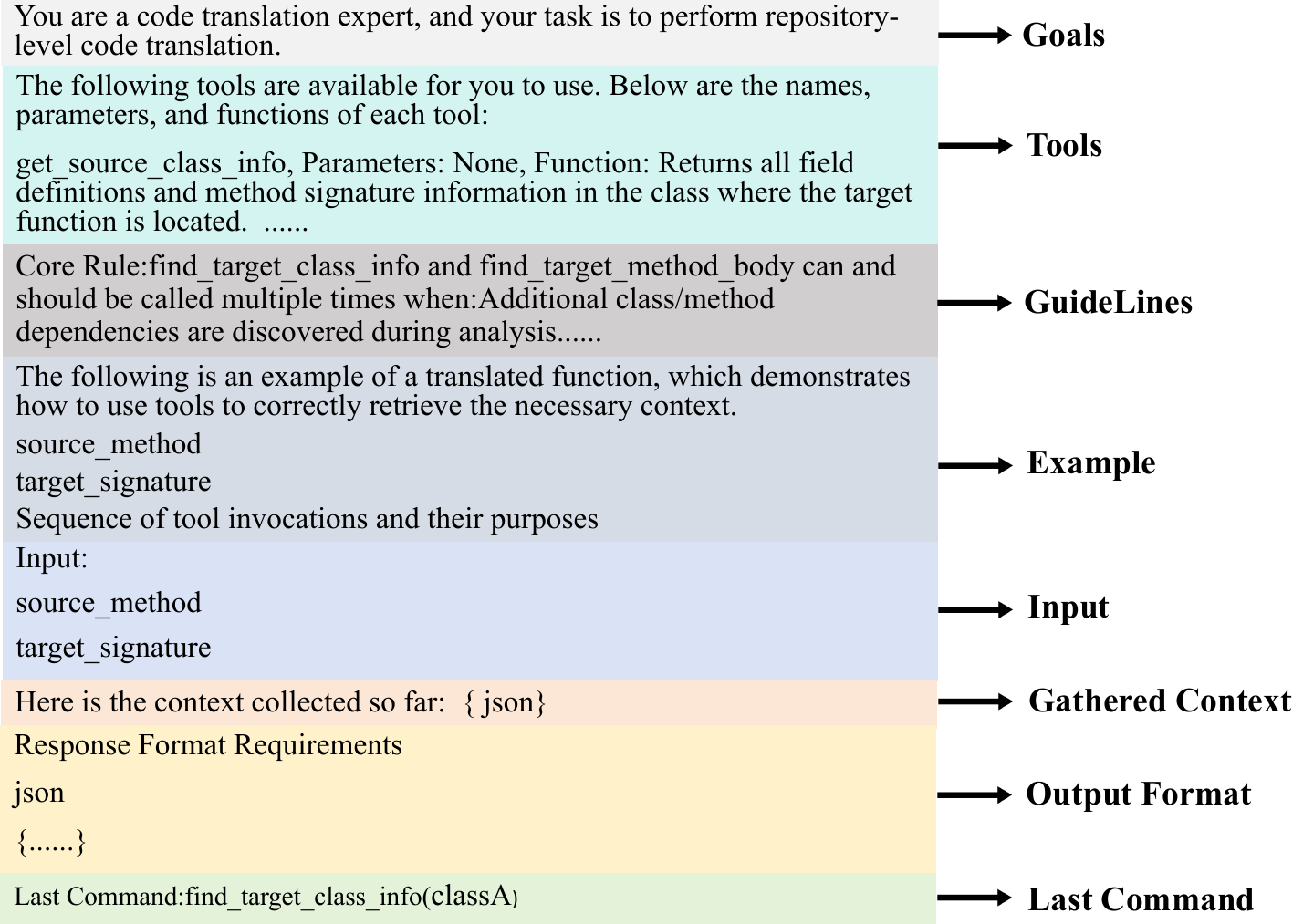}}
\caption{An example of the dynamic prompt construction process}
\label{fig:dynamic_prompt}
\end{figure}

To enable the agent to accurately invoke the necessary tools, we design a dynamic prompt to query the agent. This prompt is composed of a series of static and dynamic components. Fig.~\ref{fig:dynamic_prompt} illustrates the construction process of our dynamic prompt.

\paragraph{Goals (Static)} This section defines the agent's area of expertise as repository-aware code translation. It clarifies that the agent's objective is to collect the most relevant context to assist in performing repository-aware code translation.

\paragraph{Tools (Static)} In this section, we introduce all available tools, including their names, parameters, and descriptions. This serves as a reference for the agent, providing the basic information needed to correctly invoke and utilize each tool.

\paragraph{Guidelines (Static)} We provide a set of guidelines to clarify the fundamental principles for using the tools. First, the agent is instructed to prioritize calling the three tools: get\_source\_class\_info, get\_target\_class\_info, and find\_target\_imports, in order to obtain basic information relevant to translating the source function. Then, we demonstrate how to use the find\_target\_class\_info and find\_target\_method\_body tools. The agent is expected to repeatedly use find\_target\_class\_info and find\_target\_method\_body to gather sufficient context, including identifying new type dependencies, verifying interface implementations, and resolving mappings from packages to namespaces. Additionally, the agent is reminded to pay close attention to language-specific differences.

\paragraph{Example (Static)} We provide an example that includes the signatures of a source function and its corresponding target function to demonstrate how to properly use the tools to collect the necessary context for translating the source function. This example outlines each step of tool usage along with the rationale behind it, showing the agent how to progressively and effectively gather sufficient context.

\paragraph{Input (Dynamic)} This section includes the source function to be translated and the signature of the target function. It serves as the foundation for the agent to gather context, guiding the agent to collect context specifically related to the source function.

\paragraph{Gathered Context (Dynamic)} In this section, we list all the contextual information previously collected through tool invocations by the agent (this component is omitted during the first query to the agent). It acts as the agent's memory, allowing it to determine the next appropriate tool to call based on the already gathered context, enabling a step-by-step refinement to retrieve more relevant information.

\paragraph{Output Format (Static)} This section specifies the required response format for the agent, mandating that all responses strictly follow the JSON format and include three fields: id, name, and args. We explicitly instruct the agent not to include any content outside of the JSON object, such as additional text or code, and provide an example to illustrate the correct response format.

\paragraph{Last Command (Dynamic)} This section records the tool invoked in the agent's previous action along with its returned output. It serves to remind the agent to reflect on whether its current strategy may be flawed and to avoid redundant tool invocations.

\subsection{Refine Agent}
The Refine Agent serves as the core component responsible for code translation and iterative error correction through a systematic refinement process. It leverages test case execution feedback and self-reflection mechanisms to progressively enhance translation quality. The Refine Agent operates through four sequential phases: Initial Generation, Test Execution, Reflection, and Error Correction.

\subsubsection{Initial Generation}
In the initial generation phase, the Refine Agent is configure with specialized expertise in repository-aware code translation. The agent receives the source function body and target function signature as primary inputs, supplemented by contextual information from the Context Agent and similar functions from the RAG Agent.

The agent is instructed to leverage this comprehensive contextual information to produce accurate translations that account for repository-specific differences. We provide detailed guidelines and concrete examples demonstrating how to effectively utilize the additional information to resolve discrepancies between source and target repositories. The agent outputs the translated code in markdown format to ensure consistent extraction and processing.

\subsubsection{Test Execution}
Following code generation, the translated function undergoes validation through the target repository's test suite. If all test cases pass successfully, the translation is deemed complete and the process terminates. Otherwise, we systematically collect and analyze test execution results following established methodologies from existing approaches.

We categorize execution failures into four distinct types: compilation errors, test failures, runtime errors, and non-terminating executions. For each error category, we extract comprehensive diagnostic information including error logs, error classifications, output messages, and precise error locations to facilitate targeted error correction in subsequent iterations.

\subsubsection{Reflection}
The reflection phase implements a self-analysis mechanism where the agent examines the relationship between the source function, target function signature, the generated erroneous code, and the corresponding test execution results. Through this systematic analysis, the agent identifies the root causes of translation failures and formulates specific correction strategies.

This reflective process enables the agent to understand not only what went wrong but also why the error occurred, leading to more informed and targeted corrections in the subsequent iteration.

\subsubsection{Error Correction}
During the error correction phase, the agent generates an improved version of the translated code by incorporating insights from the previous iteration's code, test execution results, and reflection analysis. The prompt structure mirrors that of the initial generation phase but is enhanced with three critical components: the erroneous code from the previous iteration, detailed test execution results, and the agent's reflection and correction strategy.

The corrected code undergoes immediate re-evaluation through the test suite, initiating a new cycle of Test Execution and Reflection if errors persist. This iterative process continues until either all test cases pass successfully or the predefined maximum iteration limit is reached, ensuring both translation quality and computational efficiency.

\label{sec:approach}

\section{Experiment}
This section presents the experimental design and methodology employ to evaluate \toolname. We first describe the constructed datasets, followed by an introduction to the baseline approaches. Subsequently, we detail the performance metrics and experimental configuration utilized in our evaluation.

\subsection{Dataset}
\label{sec:dataset}
We construct a comprehensive dataset to evaluate \toolname, comprising six highly-rated GitHub projects, each featuring both Java and C\# implementations. Following the established methodology of Methods2Test~\cite{tufano2022methods2test}, we extract Java-C\# translation pairs that include corresponding test cases. All unit tests within each project are executed to ensure their functionality and correctness. Detailed statistics for each project are provided in Table~\ref{tab:dataset}.

\begin{table}[htbp]
  \centering
  \caption{The statistics of the constructed dataset}
  \resizebox{.65\linewidth}{!}{%
    \begin{tabular}{l|cc} 
    \toprule
    \textbf{Project} & \textbf{Star} & \makecell{\textbf{\# Focal Method}\\\textbf{(Java / C\#)}} \\ 
    \midrule
    \textbf{\href{https://github.com/apache/lucene}{lucene}} & 2.9k & 113/114 \\
    \textbf{\href{https://github.com/apache/poi}{poi}} & 2.0k & 229/291 \\ 
    \textbf{\href{https://github.com/eclipse-jgit/jgit}{jgit}} & 0.2k & 134/140 \\ 
    \textbf{\href{https://github.com/itext/itext-java}{itext}} & 2.1k & 67/35 \\ 
    \textbf{\href{https://github.com/quartz-scheduler/quartz}{quartz}} & 6.5k & 42/25 \\ 
    \textbf{\href{https://github.com/apache/rocketmq-clients}{rocketmq-clients}} & 0.4k & 42/50 \\ 
    \bottomrule
    \end{tabular}
  }
  \label{tab:dataset}%
\end{table}%

\subsection{Baselines}
To evaluate the effectiveness of \toolname, we selected UniTrans~\cite{yang2024exploring} and PLTranslation~\cite{pan2024lost} as comparison benchmarks. Both UniTrans and PLTranslation represent traditional LLM-based code translation approaches that utilize test cases to validate translated code and iteratively refine translations based on test feedback. Additionally, we include the No Agent approach as a baseline, wherein the LLM directly translates the source function without any iterative refinement process.

\subsection{Evaluation Metrics}
Consistent with previous research~\cite{yang2024exploring,pan2024lost,yin2024rectifier}, we employ compile rate and pass rate as evaluation metrics. A piece of code is considered compilable if it successfully passes the compiler's syntactic and semantic checks, and correct if it additionally passes all associated test cases. The compile rate and pass rate are formally defined as the proportion of all generated functions that are compilable and functionally correct, respectively.

\subsection{Experimental Settings}
We implemented \toolname using Python, leveraging the LangChain API to interface with the relevant models, including Llama 3.1 8B and 70B~\cite{grattafiori2024llama}, Qwen2.5 7B and 72B~\cite{yang2025qwen3}, GPT4o-mini~\cite{hurst2024gpt}, and DeepSeek V3~\cite{liu2024deepseek}. We adhere to established best practice guidelines in designing our prompts. For baseline comparisons, we adapt their prompts to ensure applicability to repository-aware translation scenarios and guarantee fair performance evaluation.
\label{sec:experiment}

\begin{table*}[!htbp]
\centering
\caption{RQ-1: \toolname vs. Baselines across different projects in compile rate and pass rate (C\# to Java / Java to C\#)}
\resizebox{\linewidth}{!}{
\begin{tabular}{l|cccc|cccc}
\toprule
\multirow{2}{*}{Projects} & \multicolumn{4}{c|}{Compile Rate (C\#→Java / Java→C\#)} & \multicolumn{4}{c}{Pass Rate (C\#→Java / Java→C\#)} \\
\cmidrule{2-9}
& No Agent & \toolname & UniTrans & PLTranslation & No Agent & \toolname & UniTrans & PLTranslation \\ 
\midrule
lucene & 16.81\% / 4.39\% & 49.56\% / 15.79\% & 24.78\% / 3.51\% & 24.78\% / 8.77\% & 11.50\% / 3.51\% & 36.29\% / 11.40\% & 14.16\% / 2.63\% & 21.24\% / 7.02\% \\
poi & 39.30\% / 30.58\% & 71.62\% / 66.32\% & 42.79\% / 36.43\% & 46.72\% / 63.57\% & 31.00\% / 29.55\% & 58.52\% / 57.73\% & 37.07\% / 30.58\% & 43.23\% / 36.08\% \\
jgit & 17.16\% / 11.43\% & 42.54\% / 32.86\% & 14.18\% / 16.43\% & 20.15\% / 18.57\% & 9.70\% / 7.86\% & 35.82\% / 20.71\% & 7.46\% / 10.71\% & 15.67\% / 11.43\% \\
itext & 28.36\% / 25.71\% & 61.19\% / 71.43\% & 43.28\% / 11.43\% & 38.81\% / 14.29\% & 28.36\% / 5.71\% & 61.19\% / 54.29\% & 43.28\% / 5.71\% & 38.81\% / 5.71\% \\
quartz & 54.76\% / 8.00\% & 73.81\% / 48.00\% & 54.76\% / 8.00\% & 50.00\% / 8.00\% & 30.95\% / 8.00\% & 57.14\% / 36.00\% & 47.62\% / 8.00\% & 47.62\% / 8.00\% \\
rocketmq-clients & 0.00\% / 4.00\% & 33.33\% / 24.00\% & 2.38\% / 10.00\% & 2.38\% / 4.00\% & 0.00\% / 2.00\% & 26.19\% / 14.00\% & 2.38\% / 8.00\% & 2.38\% / 4.00\% \\
\midrule
Average & 26.07\% / 14.03\% & 55.34\% / 43.07\% & 30.36\% / 14.30\% & 30.47\% / 19.53\% & 18.59\% / 9.44\% & 45.84\% / 32.36\% & 25.33\% / 10.94\% & 28.16\% / 12.04\% \\
\bottomrule
\end{tabular}
}
\label{tab:rq1_combined}
\end{table*}

\section{Results and Analysis}

To comprehensively evaluate the effectiveness of \toolname in repository-aware code translation tasks, we formulate three research questions:

\textbf{RQ-1 Effectiveness Comparison.} How does RepoTrans-Agent perform compared to existing baseline methods in repository-aware code translation?

\textbf{RQ-2 Ablation Study.} What is the contribution of each component to the overall effectiveness of \toolname?

\textbf{RQ-3 Model-Agnostic Analysis.} 
How does RepoTrans-Agent perform across different LLM configurations?

The detailed experimental results and analysis for each research question are presented in the following subsections:

\subsection{RQ-1: Effectiveness Comparison}
\label{sec:rq1}

\noindent\textbf{Experimental Design.} To investigate whether \toolname outperforms existing methods on repository-aware code translation tasks, we conducted comprehensive experiments on our constructed dataset. For each source function in the dataset, we performed translation using four distinct methods and determined whether the translated function succeeded by executing the project's test cases. We uniformly employed the DeepSeek V3 model for code translation and maintained consistent model parameters across all methods to eliminate confounding factors. Finally, we assessed the effectiveness of each method using compile rate and pass rate as evaluation metrics.

\noindent\textbf{Results.} The experimental results encompass four components: Overall Results, Statistics of Error Types, Comparison with Baselines, and Case Study.

\paragraph{Overall Results} Table~\ref{tab:rq1_combined} presents the compile rate and pass rate achieved when employing the DeepSeek V3 model. \toolname and the three baseline approaches were evaluated on our dataset for bidirectional translation between Java and C\#. For the 627 functions translated from C\# to Java, our method successfully compiled 363 functions and 299 functions passed all test cases. For the 655 functions translated from Java to C\#, 306 functions were successfully compiled and 245 functions passed all test cases.

\paragraph{Statistics of Error Types} For functions that failed to pass all test cases, we extracted the error logs and systematically categorized the error types. The results are presented in Table~\ref{locationTypeResults}. 

The predominant errors were compilation errors, with the most frequent issue being ``symbol not found''. 
This typically occurs when classes, methods, or fields defined in the source repository are absent in the target repository. This observation underscores that implementation differences between source and target repositories constitute the primary cause of translation failures, thereby emphasizing the critical importance of context collection in repository-aware code translation.

\begin{table}[!htbp]
\centering
\small
\caption{The error types of the functions that failed to be translated.}
\label{locationTypeResults}
\begin{tabular}{lcccc} 
\toprule
Error type    & C\# to Java / Java to C\#   \\ 
\midrule
Compilation Errors    & 264/349             \\ 
Runtime Errors   & 26/19                  \\ 
Functional Errors     & 36/41                   \\ 
Non-terminating Execution & 2/1             \\
\bottomrule
\end{tabular}
\end{table}

\textbf{\begin{figure}[htbp]
\centerline{\includegraphics[width=1\linewidth]{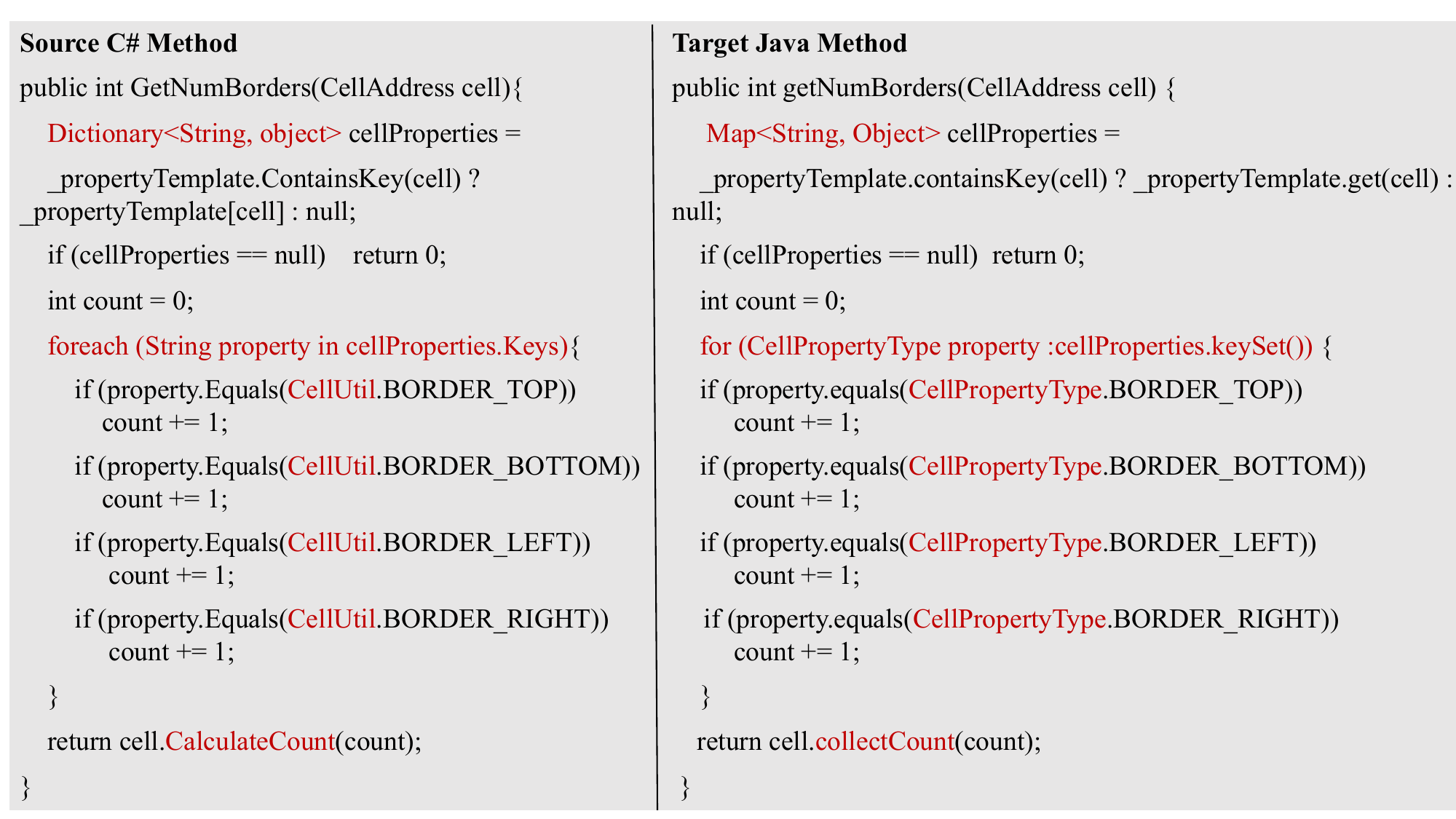}}
\caption{An example of a function translation using tools by \toolname}
\label{fig:example}
\end{figure}}

\begin{figure}[htbp]
\centerline{\includegraphics[width=0.6\linewidth]{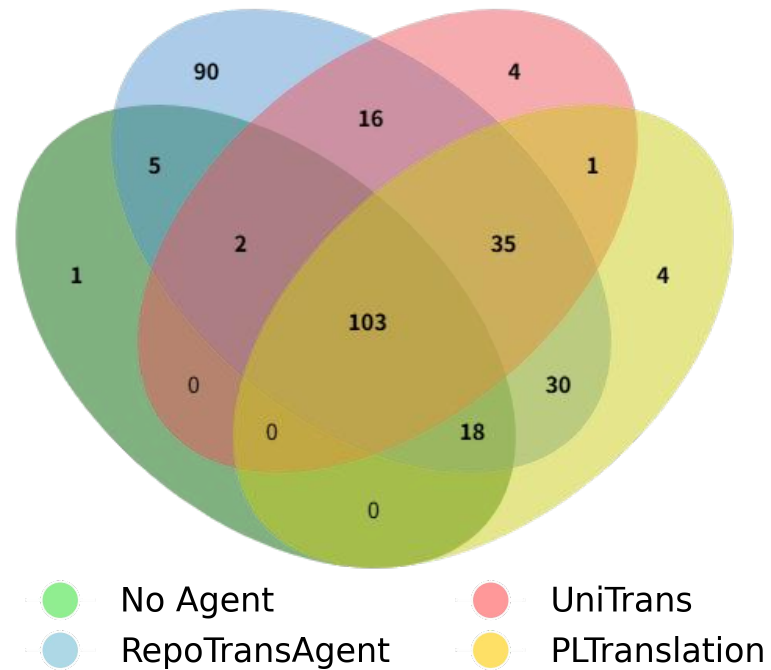}}
\caption{The intersection of functions successfully translated by different methods}
\label{fig:venn}
\end{figure}

\paragraph{Comparison with Baselines} When compared with baseline approaches, \toolname achieves substantially higher compile and pass rates across both translation directions and all projects, although the magnitude of improvement varies. This demonstrates that \toolname exhibits excellent generalization capabilities across diverse projects and demonstrates strong robustness and adaptability. Notably, compared with direct code translation methods that do not incorporate agent-based approaches, our method shows substantial performance gains, confirming its effectiveness. In particular, \toolname achieves several-fold improvements in pass rate on the jgit and rocketmq-clients projects. For further analysis, taking the translation from C\# to Java as an example, Fig.~\ref{fig:venn} illustrates the overlap of successfully translated functions among different methods.

The results demonstrate that \toolname successfully translated nearly all functions that were handled by any of the three baseline methods, with only 10 functions successfully translated by at least one baseline but not by \toolname. Moreover, \toolname successfully translated 90 functions that none of the baseline methods managed to translate. We attribute this superior performance to \toolname's capability to effectively collect the most relevant context through collaborative coordination among different agents, and to iteratively correct errors during the translation process.

\paragraph{Case Study} Fig.~\ref{fig:example} presents a function that was successfully translated exclusively by \toolname. During the translation process, the agent first adhered to established guidelines, converting the C\# Dictionary data structure into Java's Map, and translating the C\# foreach loop into a Java for loop. Subsequently, the agent invoked the find\_target\_imports tool and discovered that the target file did not import the CellUtil enum, but instead imported the CellPropertyType enum. Consequently, the agent appropriately translated CellUtil into CellPropertyType.

Subsequently, by invoking the find\_target\_class\_info tool, the agent identified that the CellAddress class in the target repository did not contain the calculateCount method but did include a method named collectCount. Through semantic inference from the method name that the two methods were likely functionally equivalent, the agent replaced cell.CalculateCount with cell.collectCount in the translation.

Through these systematic steps, the agent successfully translated the source function. This example demonstrates the effectiveness of each tool function: the agent can selectively gather the context necessary for translating the source function and exhibits reasoning capabilities, enabling it to address challenges that static analysis cannot resolve.

\intuition{
{\bf Answer to RQ-1:}
\toolname significantly outperforms the baseline approaches in improving both compile and pass rates, with the average compile rate increasing from 26.07\%–30.47\% to 55.34\%, and the average pass rate rising from 18.59\%–28.16\% to 45.84\%. Furthermore, \toolname demonstrates consistent performance across various projects, exhibiting strong generalization capabilities.    
}

\subsection{RQ-2: Ablation Study}
\label{sec:rq2}

\begin{figure*}[htbp]
\centerline{\includegraphics[width=.85\linewidth]{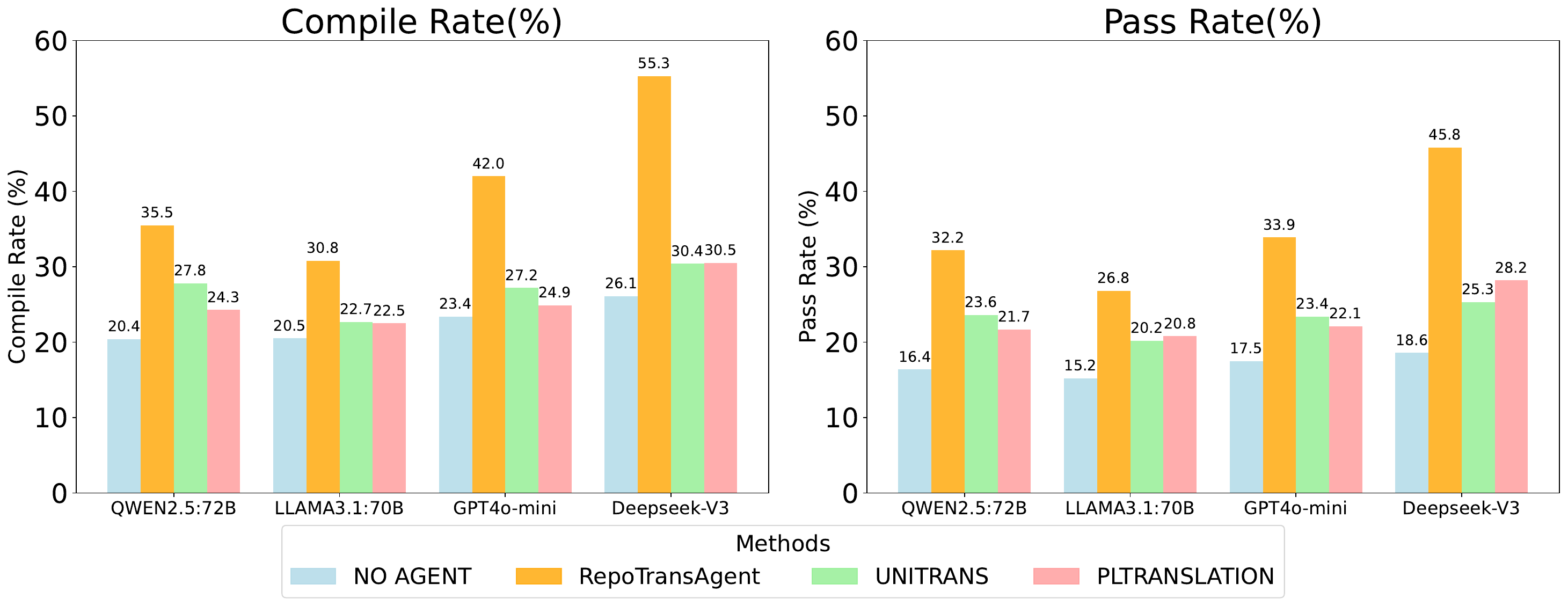}}
\caption{\toolname vs. Baselines across different LLMs on average compile rate and average pass rate}
\label{fig:RQ3.1}
\end{figure*}

\begin{figure*}[htbp]
\centerline{\includegraphics[width=.85\linewidth]{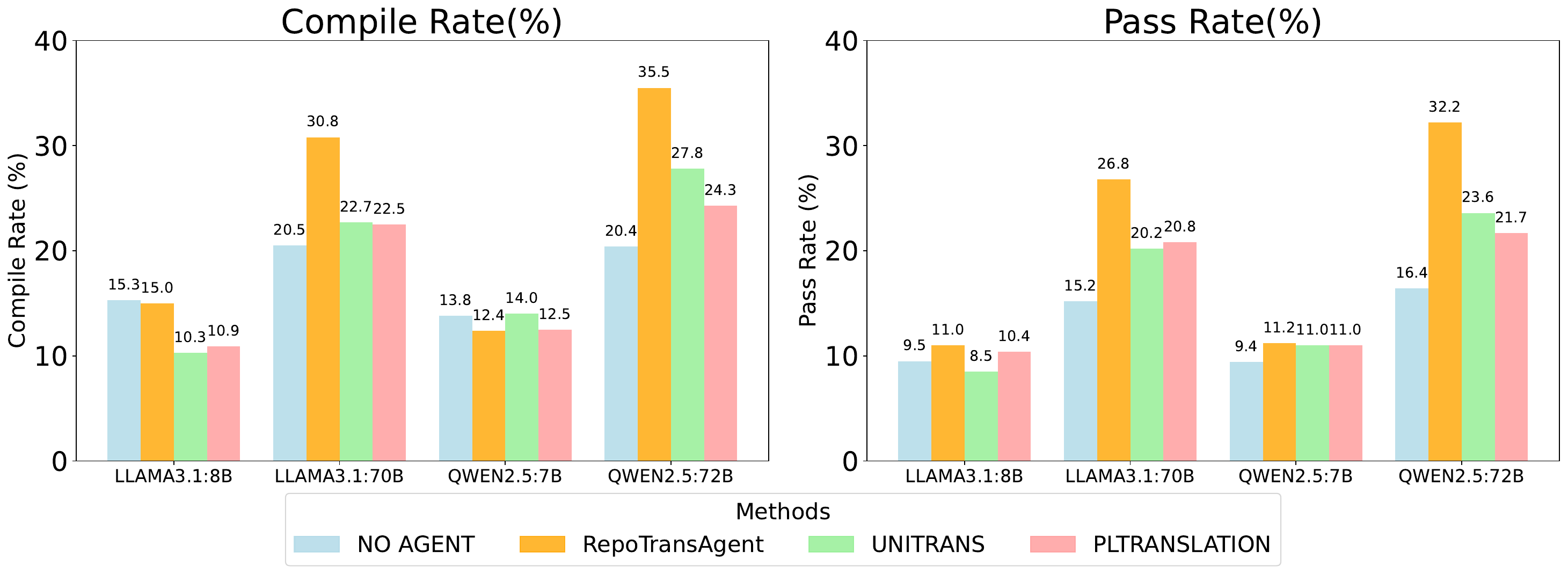}}
\caption{Average compile rate and average pass rate of LLMs with different sizes}
\label{fig:RQ.3.2}
\end{figure*}

\noindent\textbf{Experimental Design.} To comprehensively understand the contribution of each agent in \toolname to translation performance, we conducted a systematic ablation study. We employed the DeepSeek V3 model and evaluated six projects involving translation from C\# to Java. When the RAG Agent was removed, similar functions were not provided to the Refine Agent; when the Context Agent was removed, contextual information was absent from the input to the Refine Agent; and when the Refine Agent was removed, only single-pass generation was performed without iterative error correction. After removing each corresponding agent, we utilized the remaining components for code translation and continued to assess performance based on compile rate and pass rate.

\begin{table}[htbp]
\centering
\caption{Different configurations of \toolname}
\label{tab:ablation}
\resizebox{.85\linewidth}{!}
{
    \begin{tabular}{lcc} 
    \toprule
    Project         & Compile Rate & Pass Rate \\
    \midrule
    No RAG Agent & 51.39\%        & 42.94\%      \\
    No Context Agent & 36.87\%        & 32.20\%      \\
    No Refine Agent & 34.81\%      & 26.50\%       \\
    \midrule
    \toolname (default)  & 55.34\%        & 45.84\%      \\
    No Agent   & 26.07\%        & 18.59\%      \\
    \bottomrule
    \end{tabular}
}
\end{table}

\noindent\textbf{Results.} The experimental results are presented in Table~\ref{tab:ablation}. Without the RAG Agent, both the compile rate and the pass rate exhibited decreases. The RAG Agent demonstrated particular effectiveness when the source function had overloaded variants, as these overloaded functions possessed highly similar structures and functionalities. However, in most cases, it proved challenging to retrieve functions in the target repository that were sufficiently similar to the source function, thus limiting the effectiveness of the RAG Agent. 

When the Context Agent and the Refine Agent were removed, there was a substantial decline in both compile rate and pass rate. Without the Context Agent, due to the absence of effective contextual information, the agent tended to directly replicate the implementation of the source function and disregard the implementation differences between source and target repositories, resulting in compilation errors. When the Refine Agent was absent, the agent could not perform error correction based on test case execution results. Due to the inherent limitations of the agent's capacity, it was challenging to generate a completely correct function in a single attempt, which led to a significant decrease in both compile rate and pass rate.

\intuition{
{\bf Answer to RQ-2:}
The ablation study demonstrates that the Context Agent and Refine Agent play crucial roles in the performance of \toolname, as the absence of either agent results in a substantial decline in both compile rate and pass rate. The contribution of the RAG Agent to the overall performance of \toolname is relatively modest.
}

\subsection{RQ-3: Model-Agnostic Analysis}
\label{sec:rq3}

\noindent\textbf{Experimental Design.} To investigate the generalization capability of \toolname across different models, we selected four representative models—Llama 3.1-70B, Qwen2.5-72B, GPT-4o-mini, and DeepSeek V3—and evaluated them on the task of translating code from C\# to Java. Additionally, to examine the impact of model size on code translation performance, we selected two size variants of both Llama 3.1 (8B and 70B) and Qwen2.5 (7B and 72B) and conducted experiments on the same translation task. We computed the average compile rate and average pass rate across all projects as evaluation metrics.

\noindent\textbf{Results.} The experimental results encompass three components: Overall Results, The Impact of Model Size on Performance, and Complementarity Among Different LLMs.

\paragraph{Overall Results} The results are presented in Fig.~\ref{fig:RQ3.1}. The findings demonstrate that \toolname substantially enhances the performance of these LLMs on repository-aware code translation tasks. For instance, on the Qwen2.5-72B model, \toolname increases the average compile rate from 20.4\%–27.8\% to 35.5\% and the average pass rate from 16.4\%–23.6\% to 32.2\%. Among the evaluated models, \toolname achieves the most substantial improvements on GPT-4o-mini and DeepSeek V3. Overall, \toolname consistently improves the compile rate and pass rate of the base LLMs across all projects. This demonstrates the effectiveness of the \toolname approach and underscores its broad applicability. \toolname is designed to be model-agnostic, enabling compatibility with various LLMs.

\paragraph{The Impact of Model Size on Performance.} The results are illustrated in Fig.~\ref{fig:RQ.3.2}. The findings reveal that \toolname achieved substantial improvements over the baseline approaches on larger models—Llama 3.1-70B and Qwen2.5-72B. However, the improvements were considerably smaller on smaller models: Llama 3.1-8B and Qwen2.5-7B, and even exhibited decline in some cases. This indicates that the performance of \toolname is significantly influenced by model size. The primary reasons are that smaller models struggle to invoke tool functions effectively, resulting in suboptimal context collection by the Context Agent. Additionally, due to the relatively extensive prompts designed for our method, smaller models experience difficulty handling long contexts and often fail to capture critical portions of the instructions.

\begin{figure}[htbp]
\centerline{\includegraphics[width=0.6\linewidth]{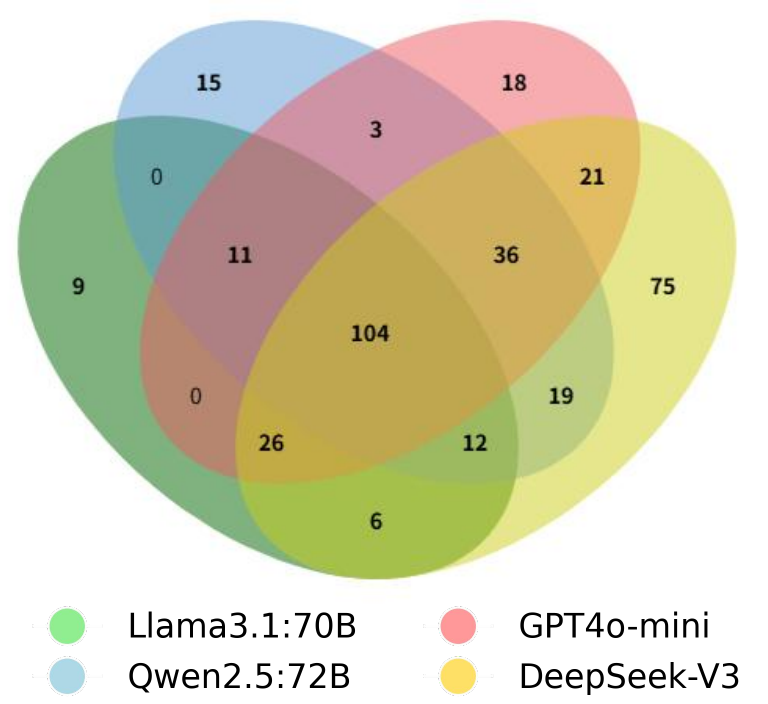}}
\caption{The intersection of functions successfully translated by
different LLMs}
\label{fig:venn_2}
\end{figure}

\paragraph{Complementarity Among Different LLMs.} To further analyze the sets of functions successfully translated by different models, Fig.~\ref{fig:venn_2} presents the overlap among them. The results reveal that the models exhibit complementarity: 104 functions were successfully translated by all four models, while a total of 355 functions were successfully translated when combining the outputs of all models. This indicates that leveraging multiple models can achieve a higher number of successfully translated functions.

\intuition{
{\bf Answer to RQ-3:}
\toolname substantially enhances the performance of base LLMs on repository-aware code translation tasks through its multi-agent framework. Additionally, the functions successfully translated by different models demonstrate complementarity, and \toolname exhibits superior performance with larger models compared to smaller ones.
}

\label{sec:results}

\section{Threats to Validity}

\noindent\textbf{Internal Validity.} A primary internal threat concerns potential data leakage, as some evaluated functions may appear in the training data of LLMs such as DeepSeek V3. To address this concern, we perform exact matching between the translated functions and ground-truth implementations after removing line breaks, comments, and extraneous whitespace. Among 299 C\#→Java and 245 Java→C\# translations, only 19.40\% and 13.06\% constitute exact matches, respectively, indicating a low likelihood of data leakage. Additionally, we ensure that all unit tests are executed to verify their functionality, further supporting the reliability of our evaluation. Another internal threat pertains to the stochasticity of LLM-based agents, which may lead to output variance. To mitigate this issue, we configure the temperature of the RAG Agent and Context Agent to 0, minimizing randomness, while the Refine Agent employs a temperature of 0.8 to encourage diverse hypotheses.

\noindent\textbf{External Validity.} Our evaluation is currently constrained to Java and C\# projects, which may limit the generalizability of \toolname to other programming languages. To enhance external validity, future work will extend the evaluation to additional programming languages and more diverse codebases.


\section{Related Work}
\label{sec:related}

\subsection{Code Translation}
Code translation facilitates cross-platform interoperability and promotes code reuse across diverse programming environments, playing a crucial role in legacy system modernization~\cite{lu2021codexglue,puri2021codenet}.

Early approaches evolved from rule-based systems with handcrafted transformation rules to Statistical Machine Translation (SMT) for modeling common code patterns~\cite{nguyen2013lexical,chen2018tree}, and subsequently to neural network models, though early RNNs struggled with complex syntactic structures~\cite{dong2016language,yin2017syntactic,yin2018tranx}.

LLMs have revolutionized code translation with unprecedented capabilities, though challenges remain in translation accuracy and repository-specific context handling~\cite{zhu2022xlcost,rithy2022xtest}.
Recent work has addressed these challenges through various approaches: benchmarking LLM limitations in complex scenarios~\cite{jiao2023evaluation}, context-enriched prompts~\cite{pan2024lost}, test case feedback for iterative refinement~\cite{yang2024exploring}, consistent prompt design~\cite{macedo2024exploring}, and comprehensive repository-aware evaluation~\cite{ou2024repository}. Despite these advances, repository-aware contextual understanding remains largely unexplored. Our work addresses this limitation through a multi-agent framework specifically designed for repository-aware code translation.

\subsection{LLM-based Multi-Agent Systems}
LLM-based multi-agent systems have emerged as a promising research direction, advancing several areas. 
Infrastructure development focuses on frameworks for efficient agent coordination and task management~\cite{gong2023mindagent,chen2023agentverse,zhang2023building,hong2023metagpt,zhang2024proagent}. 
Benchmark development evaluates multi-agent performance in dynamic environments~\cite{chen2024llmarena,dong2024villageragent}. 
Large-scale social simulations explore modeling complex societal behaviors~\cite{al2024project,li2022evolution,de2004epidemiology}. 
Domain-specific applications demonstrate practical effectiveness in specialized scenarios, including software engineering tasks~\cite{d2024marg,zhang2024autocoderover,yang2024swe}.
Our work contributes to this research by introducing a specialized multi-agent framework for repository-aware code translation. Unlike general-purpose systems, \toolname leverages domain-specific knowledge and specialized agents to address the unique challenges of cross-language code translation in real-world software repositories.
\label{sec:related_work}

\section{Conclusion and Future Work}

We present \toolname, a novel multi-agent LLM framework designed for repository-aware code translation. \toolname systematically decomposes the translation process into specialized subtasks, including context retrieval, dynamic prompt construction, and iterative code refinement, each managed by dedicated agents. By leveraging retrieval-augmented generation and dynamic prompting strategies, our approach effectively addresses the fundamental challenges of insufficient context awareness, rigid prompt design, and ineffective error correction mechanisms found in existing approaches.Comprehensive experiments conducted on hundreds of Java-C\# translation pairs from six open-source projects demonstrate that \toolname significantly outperforms state-of-the-art baseline methods in both compile rate and pass rate, achieving up to 55.34\% compilation rate and 45.84\% pass rate. Further analysis confirms the robustness and generalizability of \toolname across different LLM architectures, underscoring its practical value for real-world repository-aware code translation applications.


\balance
\bibliographystyle{IEEEtran}
\bibliography{main}

\end{document}